\documentclass[a4paper,twocolumn,11pt,accepted=2020-10-02]{quantumarticle}
\pdfoutput=1
\usepackage[utf8]{inputenc}
\usepackage[english]{babel}
\usepackage[T1]{fontenc}
\usepackage{hyperref}

\usepackage[pdftex]{graphicx}  
\usepackage{xcolor}
\usepackage{dcolumn}
\usepackage{bm}
\usepackage{amsmath,amsthm,amssymb,mathrsfs,mathtools,amsfonts}
\usepackage{verbatim}
\usepackage{ulem}
\usepackage{epsfig}
\usepackage{color}
\usepackage{xfrac}
\usepackage[numbers,sort&compress]{natbib}
\emergencystretch=\maxdimen

\newcommand{\be}{\begin{equation}}
\newcommand{\ee}{\end{equation}}

\newcommand{\bra}[1]{\left\langle #1 \right|}
\newcommand{\ket}[1]{\left| #1 \right\rangle}
\newcommand{\braket}[2]{\left\langle #1 \middle| #2 \right\rangle}

\newcommand{\cd}[1]{c^{\dagger}_{#1}}

\begin{document}
\title{Weak-ergodicity-breaking via lattice supersymmetry}
\author{F. M. Surace}
\affiliation{The Abdus Salam International Center for Theoretical Physics, Strada Costiera  11,  34151  Trieste, Italy, and
SISSA, via Bonomea 265, 34136 Trieste, Italy}
\author{G. Giudici}
\affiliation{The Abdus Salam International Center for Theoretical Physics, Strada Costiera  11,  34151  Trieste, Italy, and
SISSA, via Bonomea 265, 34136 Trieste, Italy}
\author{M. Dalmonte}
\affiliation{The Abdus Salam International Center for Theoretical Physics, Strada Costiera  11,  34151  Trieste, Italy, and
SISSA, via Bonomea 265, 34136 Trieste, Italy}


\begin{abstract}

We study the spectral properties of $D$-dimensional $N=2$ supersymmetric lattice models. We find systematic departures from the eigenstate thermalization hypothesis (ETH) in the form of
a degenerate set of ETH-violating supersymmetric (SUSY) doublets, also referred to as many-body scars, that we construct analytically. These states are stable against arbitrary SUSY-preserving perturbations, including  inhomogeneous couplings. For the specific case of two-leg ladders, we provide extensive numerical evidence that shows how those states are the only ones violating the ETH, and discuss their robustness to SUSY-violating perturbations. Our work suggests a generic mechanism to stabilize quantum many-body scars in lattice models in arbitrary dimensions.

\end{abstract} 

\maketitle

\section{Introduction}

In many-body theories, generic phenomena are often associated to and characterized by the presence of symmetries~\cite{coleman_1985}. Examples include quantum critical points and topological insulators~\cite{fradkinbook}, whose universal properties are dictated by the presence of microscopic global symmetries, and the confining properties of gauge theories, which are often related to the structure of local conservation laws~\cite{Creutz1997}. While these examples concern the equilibrium properties of matter, the role of symmetries has also been widely investigated in systems out-of-equilibrium, for instance, in connection to universal behavior~\cite{Polkovnikov2011a,Sieberer_2016}.

A paradigmatic phenomenon that - in a sense we specify below - lies 'in- between' equilibrium and out-of-equilibrium is represented by eigenstates of quantum Hamiltonians that, despite belonging to the middle of the energy spectrum, feature properties that are at odds with theoretical expectations based on the eigenstate thermalization hypothesis (ETH)~\cite{Deutsch1991, Srednicki1994,Brandino2012,Delfino_2014}. These states, recently dubbed quantum many-body scars~\cite{Turner2018}, have finite energy density above the ground state and sub-extensive entanglement entropy. Quantum scars have been linked to anomalously slow dynamics observed in laser-dressed Rydberg atom ensembles~\cite{Bernien2017}, and have attracted a considerable theoretical effort~\cite{Turner2018,Turner2018a,Khemani2019,Choi2019,Ho2019,LinMPS2019,Iadecola2019,Surace,Michailidis2019,Moudgalya2020}. 
While a number of models supporting quantum many-body scars have recently been found~\cite{Moudgalya2018,Moudgalya2018a,Mark2019,Schecter2019,Bull2019,Iadecola2020,Ok2019,Hudomal2019,Shibata2019,Chattopadhyay2019,Pai2019,Moudgalya2019,Mark2020,Zhao2020,Lee2020}, the general conditions (if any) for stabilizing ETH-violating states are still unknown, and the role of symmetries in this context stands as an open question.
Some of the recent works in this direction link the presence of quantum scars to signatures of integrability~\cite{Khemani2019}, to semiclassical trajectories~\cite{Ho2019,Michailidis2019}, to quasiparticle excitations \cite{LinMPS2019,Iadecola2019} and to the emergence of an algebraic structure~\cite{Choi2019}. Another candidate mechanism was put forward in Ref.~\cite{Surace}, in which scarred bands of Rydberg atom chains are interpreted as special eigenstates that survive the lattice regularization of an integrable field theory. While integrability does not have an immediate counterpart in more than one dimension, the Coleman-Mandula theorem shows how supersymmetry provides a feasible way of extending the set of conservation laws without resulting in a trivial (in the sense of S-matrix being the identity) theory~\cite{ColemanMandula}.

Here, we show how supersymmetry (SUSY) provides a route to formulate lattice models with 'scarred' states in the middle of the spectrum, whose stability is guaranteed as long as supersymmetry itself is not violated. In concrete, we consider $D$-dimensional lattice models of constrained spin-less fermions, that realize an exact $N=2$ supersymmetry at the lattice level~\cite{Fendley2003,Fendley2003a,Fendley2005,Huijse2008,Huijse2008a,Cheong2009,Beccaria2012,Bauer2013,Hagendorf2013}, and show how these models support scarred eigenstates (as SUSY doublets) in any $D$. After the general proof, we discuss in detail the ladder case, and address the resilience of scarred eigenstates in the presence of supersymmetry-breaking terms.

\section{Supersymmetric lattice models}
The model we study has been introduced in Ref.~\cite{Fendley2003}. The degrees of freedom are spinless fermions $c_{\bf r}$, with ${\bf r}$ being a site on a generic lattice, and the operators satisfy the canonical anticommutation relations $\{c_{\bf r}^\dagger, c_{\bf s}\}=\delta_{{\bf r},{\bf s}}$. The Hamiltonian can be written in terms of the supercharge operators $Q$ and $Q^\dagger$ defined as
\begin{equation}
    Q^\dagger = \sum_{\bf r} \alpha^{\mathstrut}_{\bf r} P^{\mathstrut}_{\bf r}\cd{{\bf r}}, \hspace{0.2cm}  Q = \sum_{\bf r} \alpha^{\mathstrut *}_{\bf r} P^{\mathstrut}_{\bf r} c^{\mathstrut}_{\bf r}
\end{equation}

where $\alpha_{\bf r}$ is a complex coefficient, and $P_{\bf r}$ is a projector which constrains all the neighbours of site $i$ to be unoccupied. The Hamiltonian has the form
\begin{equation}
H = \{Q^\dagger, Q\}.
\end{equation}
The supercharge operators satisfy 
\begin{equation}
    Q^2=(Q^\dagger)^2=0,\qquad [H, Q]=[H, Q^\dagger]=0.
\end{equation}
In addition to these, the model has a symmetry associated to the fermion number $F=\sum_{\bf r} P_{\bf r} \cd{{\bf r}} c_{\bf r}$, with $[F, Q^\dagger]=Q^\dagger$, $[F, Q]=-Q$.
The Hamiltonian can be explicitly rewritten as $H=H_0+V$ with
\begin{equation}
\label{eq:Ham}
    H_0 = \sum_{\langle {\bf r}, {\bf s} \rangle}\left(\alpha^{\mathstrut}_{\bf r} \alpha_{\bf s}^{\mathstrut *} P^{\mathstrut}_{\bf r} \cd{{\bf r}} c^{\mathstrut}_{\bf s} P^{\mathstrut}_{\bf s}+ \text{H.c.}\right),
\end{equation}
\begin{equation}
        V = \sum_{\bf r} |\alpha_{\bf r}|^2 P_{\bf r}
\end{equation}
where $\langle {\bf r}, {\bf s}\rangle$ indicates pairs of neighbouring sites. Below, we will focus on $D$-dimensional hypercubic lattices of linear dimension $L$: some of the results are extendable to other bipartite lattices. The supersymmetric algebra imposes a specific structure of the spectrum. Eigenstates can be classified in singlets and doublets: all the singlets satisfy $Q\ket{\psi}=Q^\dagger\ket{\psi}=H\ket{\psi}=0$; doublets are pairs of states of the form $\ket{\psi}$, $Q^\dagger\ket{\psi}$ (with the condition $Q\ket{\psi}=0$) and have strictly positive energy.
As discussed in detail in Ref.~\cite{Fendley2003}, this set of constraints realizes a $N=2$ SUSY which is exact at the lattice level.

\begin{figure}[t]
\begin{center}
\includegraphics[width=0.46\textwidth]{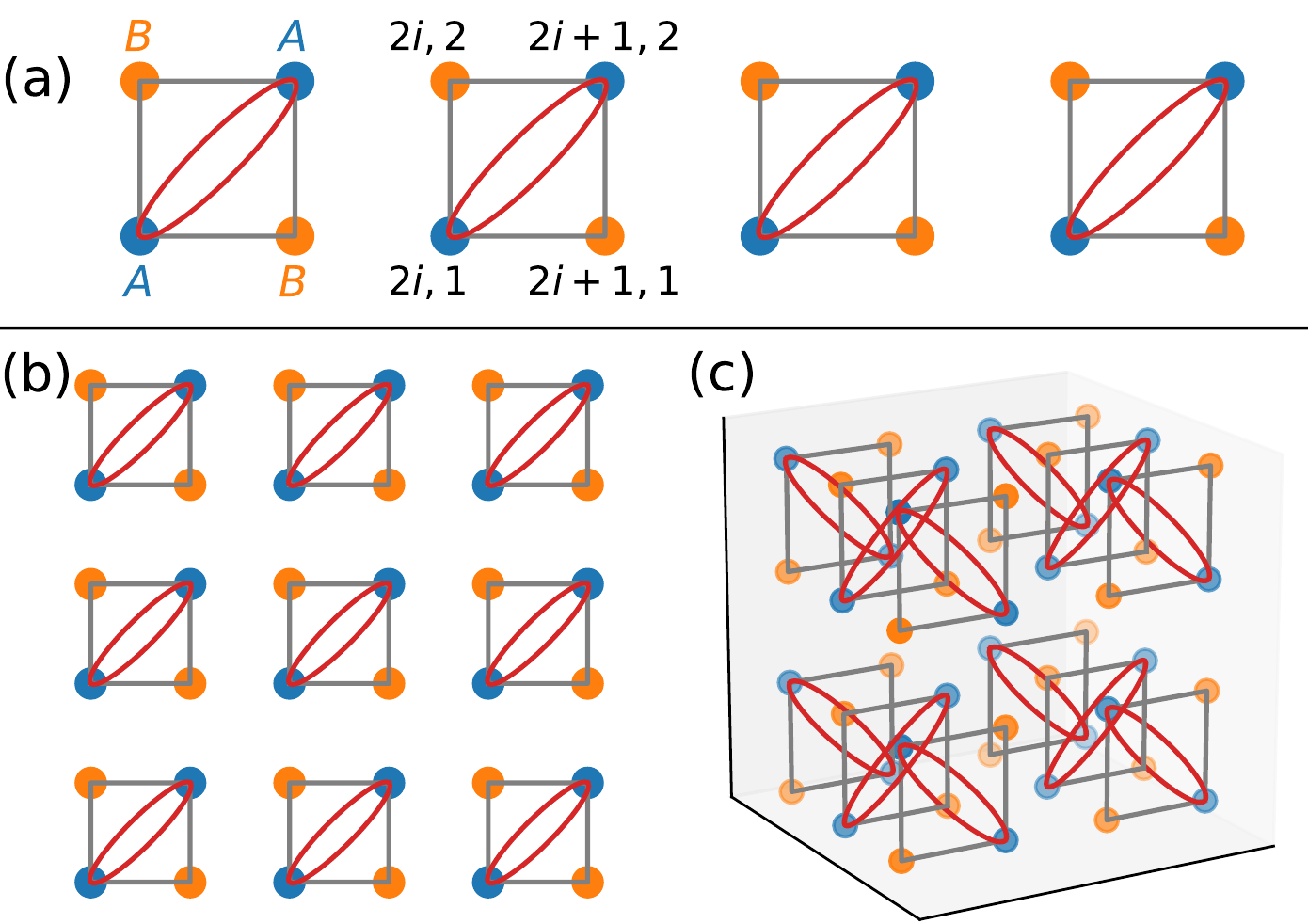}
\end{center}
\caption{(a) Exact eigenstate $\ket{\psi_{A,e}}$ in two-leg ladders. Sites belonging to sublattices $A$ and $B$ are colored in blue and orange respectively. Grey rectangles indicate the plaquettes, and for each plaquette there is a fermion in a superposition of the sites on the diagonal as in Eq.~\ref{eq:exstA}. The square (b) and cubic (c) lattices are split into plaquettes. On each plaquette we put a fermion in a superposition between the $A$ sites, in such a way that hopping terms annihilate the state. States with larger fermionic numbers can be constructed by placing two fermions, one on each $A$ site.
}
\label{fig:ladder}
\end{figure}

Before moving to the core of our work, we note that, in 2D and 3D, the models presented above draw strong similarities to the dynamics of fermionic isotopes confined in optical lattices, and laser-dressed with Rydberg $s$- or $p$-states~\cite{PhysRevLett.104.223002,PhysRevLett.104.195302,PhysRevLett.105.160404,Zeiher_2016,Glaetzle:2014dq,mitra2019robust}. In particular, the tunneling dynamics subjected to the constrains discussed above has been pointed out in Ref.~\cite{PhysRevB.92.045106, PhysRevLett.111.165302}.

\section{Exact eigenstates at finite energy density: two-leg ladders}\label{sec:exeig}

We now construct exact eigenstates in the middle of the spectrum. These states can be written as product states of square plaquettes, and can be found for any filling $F\ge L/2$, as we will detail in the following subsections. For the sake of readability, we first discuss the conceptually simpler 2-leg ladder case, and then move forward to the generic bipartite lattice in $D$ spatial dimensions.

To set the notation, we define the two sublattices $A$ and $B$ as in Fig.~\ref{fig:ladder}-a, such that each $A$ site has only neighbours of type $B$ and viceversa. We split the ladder in plaquettes (the grey squares in Fig.~\ref{fig:ladder}-a): we can choose to put the plaquettes either between neighbouring even/odd or odd/even rungs. From now on, we choose to place them between even and odd rungs as in Fig.~\ref{fig:ladder}-a. From the set of states that we will construct following this choice, we can then obtain a new set of states by applying a translation of one site along the ladder (the new states will be product states of odd/even plaquettes).

{\it Half-filling. }
Since the total number of fermions $F$ is conserved, we can construct eigenstates with a fixed filling. We first consider the sector $F=L/2$. We define the states $\ket{\psi_{A, e}}$ as follows: 

\begin{equation}
\label{eq:exstA}
    \ket{\psi_{A,e}}= \prod_{i=0}^{L/2-1}\frac{1}{\mathcal{N}_{i,A}}\left(d^\dagger_{2i,1}-d^\dagger_{2i+1, 2}\right)\ket{0},
\end{equation}
where $d_{i,j}^\dagger = \alpha_{i,j}^{-1}P^{\mathstrut}_{i,j}\cd{i,j}$ and $\mathcal{N}_{i,A}$ is a normalization constant. We choose the convention that the product is ordered from left to right.
The state is constructed as a product state of plaquettes, with a fermion in each plaquette: each fermion sits in a superposition between the two sites of a diagonal (of type $A$). 

In order to prove that $\ket{\psi_{A,e}}$ is an eigenstate, it is convenient to treat separately the hopping terms within a plaquette and those between different plaquettes. Within the plaquette, the fermions can hop from sites of the sublattice $A$ to the sublattice $B$: however, the coefficients in the superposition are such that the two contributions from the $A$ sites cancel due to destructive interference for each of the $B$ sites. On the other hand, the terms between different plaquettes would bring a fermion in a site $B$ which cannot be occupied due to the hard-core constraint, and hence annihilate the state. These two arguments prove that 
$H_0\ket{\psi_{A, e}}=0$.
The interaction term can also be easily computed by noting that $P_{i,j}=0$ for sites of lattice $B$ and $P_{i,j}=1$ for those of lattice $A$. Therefore we have
\begin{equation}
  H\ket{\psi_{A, e}}=V\ket{\psi_{A, e}}=\sum_{(i,j)\in A}|\alpha_{i,j}|^2\ket{\psi_{A, e}}.
\end{equation}
We can similarly construct the state $\ket{\psi_{B,e}}$, having fermions on sublattice $B$,
\begin{equation}
\label{eq:exstB}
    \ket{\psi_{B,e}}= \prod_{i=0}^{L/2-1}\frac{1}{\mathcal{N}_{i,B}}\left(d^\dagger_{2i,2}-d^\dagger_{2i+1, 1}\right)\ket{0}.
\end{equation}
As anticipated, other two states can be obtained by applying the translation operator, namely $\ket{\psi_{A/B,o}}=T\ket{\psi_{B/A,e}}$.
We note that, while eigenstates that occupy different sublattices are orthogonal ($\braket{\psi_{A,\, \cdot}}{\psi_{B,\, \cdot}}=0$), the eigenstates defined on the same sublattice have the same energy and are not orthogonal ($\braket{\psi_{A,e}}{\psi_{A,o}}\neq 0$ and $\braket{\psi_{B,e}}{\psi_{B,o}}\neq 0$), but they are linearly independent.
These states have energy $E_{A/B} =\sum_{(i,j)\in A/B}|\alpha_{i,j}|^2$: being eigenstates at a finite energy density above the zero-energy ground state, their entanglement entropy is expected to be proportional to the volume $L$. This is not the case: when the ladder is cut in two, the entanglement entropy is either 0 (if the cut is between two plaquettes) or a finite quantity (if the cut is within a plaquette). These eigenstates satisfy an area law entanglement at a finite energy density and hence they qualify as many-body quantum scars.

{\it Above half-filling.} For number of fermions $F>L/2$, a number of exact eigenstates can be similarly constructed as a product state of plaquettes.  
We start from one of the four states $\ket{\psi_{A/B, e/o}}$, and we choose $F-L/2$ plaquettes where to increase the fermion occupancy from $1$ to $2$ fermions:  on the selected plaquettes we place fermions on both sites of the diagonal. For example, we can add a fermion to the $j$-th plaquette on top of the state $\ket{\psi_{A,e}}$ by substituting $(d^\dagger_{2j,1}-d^\dagger_{2j+1,2})/\mathcal{N}_{j,A}$ with $P^{\mathstrut}_{2j,1}\cd{2j,1}P^{\mathstrut}_{2j+1,2}\cd{2j+1,2}$ in the product in Eq.~\ref{eq:exstA}. In this way, we obtain $\binom{F-L/2}{L/2}$ states, one for each choice of the positions of the doubly occupied plaquettes.

With the same argument used for the states at filling $F=L/2$, it is possible to prove that these states are annihilated by $H_0$ and are eigenstates of $V$ with eigenvalue $\sum_{(i,j)\in A/B}|\alpha_{i,j}|^2$.

\section{Exact scars in $d$-dimensional hypercubic lattices}
We now generalize the construction of exact eigenstates for the square ladder presented above to hypercubic lattices in dimension $D$. To do so, we group all the sites of the lattice into square plaquettes, and we construct the eigenstates as product states of plaquettes. We define the two sublattices $A$ and $B$, such that neighbouring sites belong to different sublattices. We find two classes of eigenstates: in $A$-states ($B$-states) fermions occupy the sites on sublattice $A$ ($B$) only. To construct the states, on each plaquette we create either one or two fermions on the ($A/B$) diagonal, with the same operators as in the ladder. A pictorial representation of one of these states is shown in Fig.~\ref{fig:ladder} for $d=2$ and $d=3$.

The number of exact eigenstates depends on the number of ways in which the lattice sites can be grouped in square plaquettes and it grows with the system size. For example, in the specific case $d=2$ and $F=L_x L_y/4$ (where $L_x$ and $L_y$ are even and are the number of sites in the $x$ and $y$ directions), we can construct $2^{L_x/2}+2^{L_y/2}-2$ different states for each sublattice ($A$ or $B$).

\section{Spectral statistics in two-leg ladders}
In the previous section we found an extensive number of states with finite energy density and an entanglement entropy which does not depend on $L$. We now show that the rest of the spectrum for a two-leg ladder with periodic boundary conditions is compatible with ETH.

\begin{figure}[t]
\begin{center}
\includegraphics[width=0.35\textwidth]{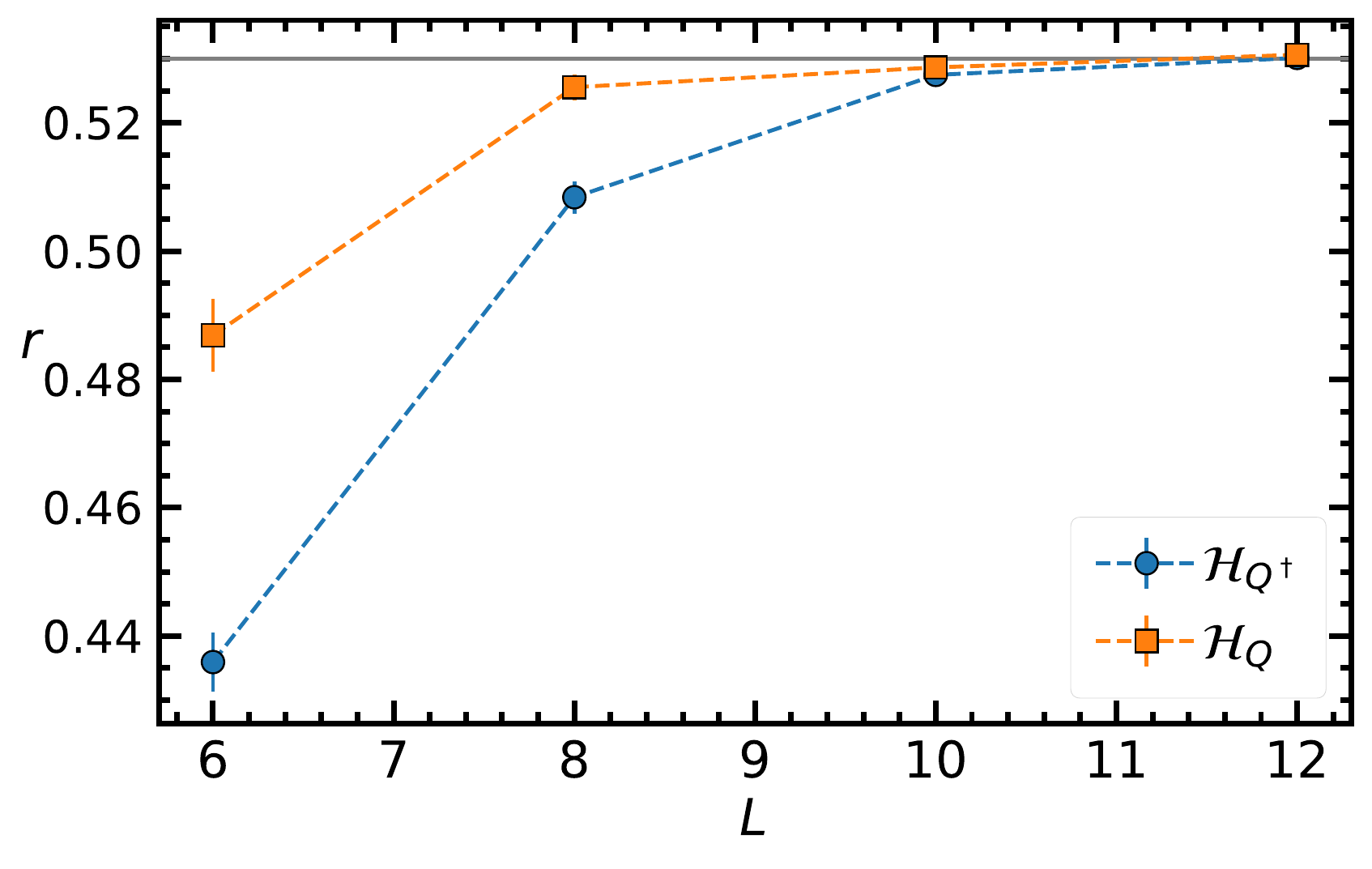}
\caption{Average level spacing ratio as a function of the number of rungs $L$ in the two sectors of non-zero energy states. The grey line indicates the value predicted for Wigner-Dyson spectral statistics. For increasing $L$, in both sectors $r$ flows towards $r_{WD}$, signalling compatibily with the ETH.
}
\label{fig:rstat}
\end{center}
\end{figure}

We study the model in Eq.~\ref{eq:Ham} using exact diagonalization. Since the construction above works for arbitrary, site-dependent coefficients $\alpha_{i,j}$, we choose random real coefficients $\alpha_{i,j}$ from a uniform distribution in the interval $[1,2)$, and average over a certain number of disorder realizations. We compute the spectrum in the sector with fermionic number $F=L/2$. Thanks to the supersymmetric algebra, the Hilbert space can be split in three sector: (i) $\mathcal{H}_{Q^\dagger} = \{\ket{\psi} : Q\ket{\psi}=0$, $Q^\dagger\ket{\psi}\neq0\}$,
(ii) $\mathcal{H}_{Q} = \{\ket{\psi} : Q\ket{\psi}\neq 0$, $Q^\dagger\ket{\psi}=0\}$, (iii) $\mathcal{H}_{0} = \{\ket{\psi} : Q\ket{\psi}= 0$, $Q^\dagger\ket{\psi}=0\}$.
The Hamiltonian is block-diagonal in these sectors: the states of the last sectors are singlets with energy $E=0$; we focus on the other two sectors, where the structure of the spectrum is non-trivial. We remark that each state of these sector belongs to a SUSY doublet and hence has a SUSY partner with the same energy, but different fermionic number ($F=L/2+1$ for the first sector and $F=L/2-1$ for the second sector). Therefore, no degeneracies and no other conservation laws are expected in the spectrum we analyze. To test the validity of the ETH for the majority of the eigenstates, we study the ratio between nearby gaps 
\begin{equation}
r_n = \frac{ \mathrm{Min} \{ \Delta E_n  , \Delta E_{n + 1 } \}  }{ \mathrm{Max} \{ \Delta E_n  , \Delta E_{n + 1 } \} }.  
\end{equation}
Here $\Delta E_n = E_{n}-E_{n-1}$, with $n$ labelling the eigenvalues $E_n$ of $H$ in increasing order, for a given disorder realization.
We then average $r_n$ over $n$ and over $100$ disorder realizations; we consider the full energy spectrum. The results, plotted in Fig.~\ref{fig:rstat} clearly show that in both sectors $r$ converges to the value expected for a Wigner-Dyson distribution $r_{WD}=0.536$ for increasing $L$, and thus validate the assumption that the majority of the eigenstates satisfy the ETH.

\begin{figure}[t]
\begin{center}
\includegraphics[width=0.48\textwidth]{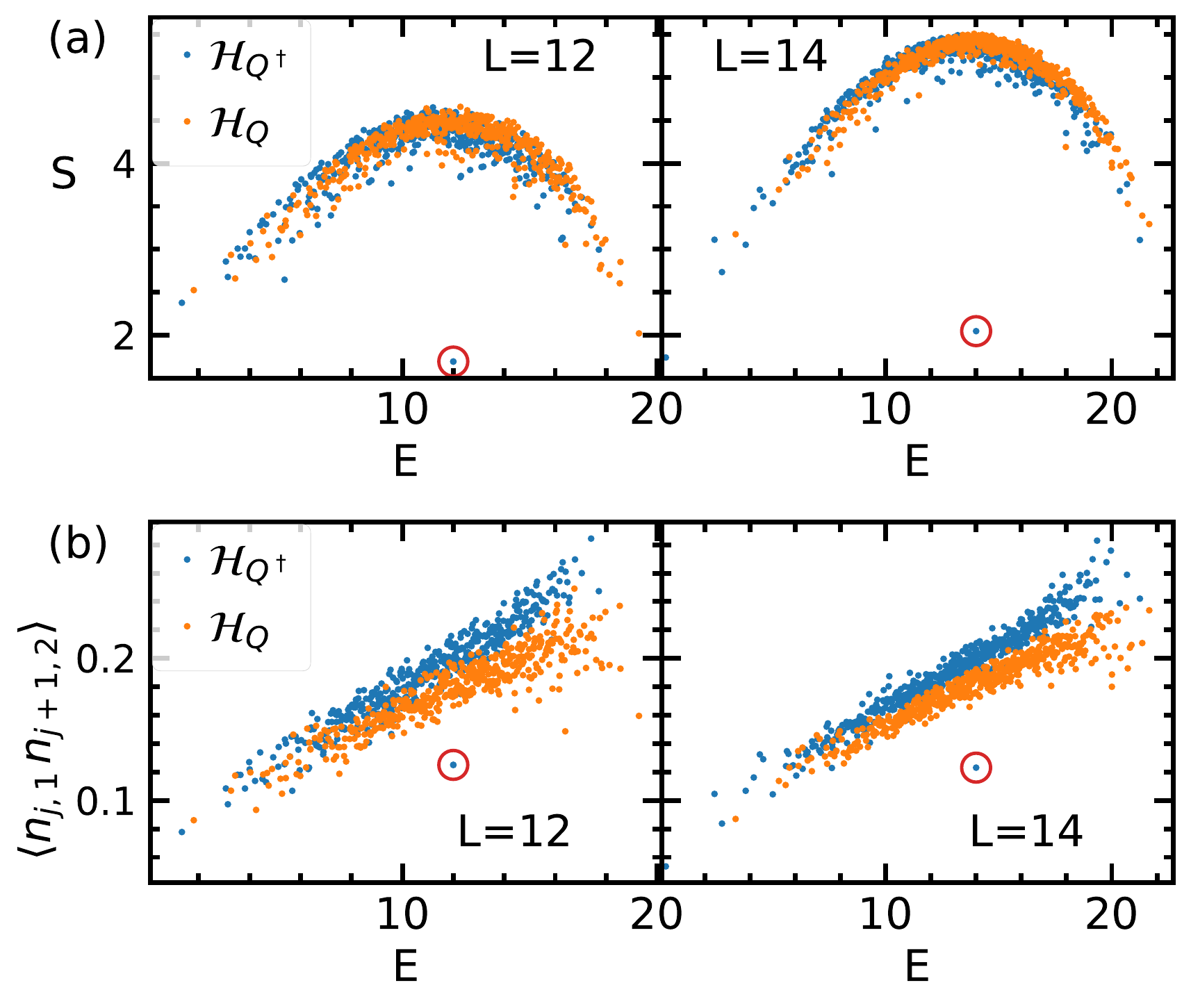}
\caption{
(a) Bipartite entanglement entropy as a function of the energy of the eigenstates in the translation- and reflection-invariant sector.
(b) Expectation value of the local observable $n_{j,1}n_{j+1,2}$ a function of the energy of the eigenstates in the translation- and reflection-invariant sector. Blue (orange) dots correspond to states in the sector $\mathcal{H}_{Q^\dagger}$ ($\mathcal{H}_{Q})$.
}
\label{fig:SvsE}
\end{center}

\end{figure}

We then check that the eigenstates we found are the only anomalous states in the spectrum. We choose coefficients $\alpha_{i,j}=1$ for all sites.
In Fig.~\ref{fig:SvsE}-a, we show the half-chain entanglement entropy for ladders of $L=12,14$ rungs with $L/2$ fermions in the translation- and reflection-invariant sector. In both sectors $\mathcal{H}_{Q^\dagger}$ and $\mathcal{H}_{Q}$, the majority of the eigenstates approximate a smooth profile with large entanglement in the middle of the spectrum, as expected in an ergodic system. A single outlier (circled in red) with anomalously small entanglement entropy is present in a region of high energy density and corresponds to the translation- and reflection-invariant superposition of the eigenstates defined above. Similar conclusions are corroborated by the analysis of diagonal correlations, depicted in Fig.~\ref{fig:SvsE}-b.

\section{Robustness to perturbations and connection to the Shiraishi-Mori construction}
We now discuss the stability of SUSY scarred eigenstates with respect to external perturbations. As discussed above, the states are stable under arbitrary supersymmetric perturbations. In particular, the construction above does not rely on any specific structure of the coefficients $\alpha_i$. In this section, we will investigate the robustness of these scarred eigenstates to other perturbations, which break the supersymmetry of the model.

As a first case, we consider the Hamiltonian  $H_{\eta}=H_0+\eta V$ 
in any $D$. If we move away from the supersymmetric point $\eta=1$, the Hamiltonian does not commute with the supercharges and the spectrum cannot be split in sectors. However, since the scars we construct (both for half and higher filling) are simultaneous eigenstates of $H_0$ and $V$, they are exact eigenstates of $H_\eta$ for arbitrary $\eta$.

\begin{figure}[t]
\begin{center}
\includegraphics[width=0.45\textwidth]{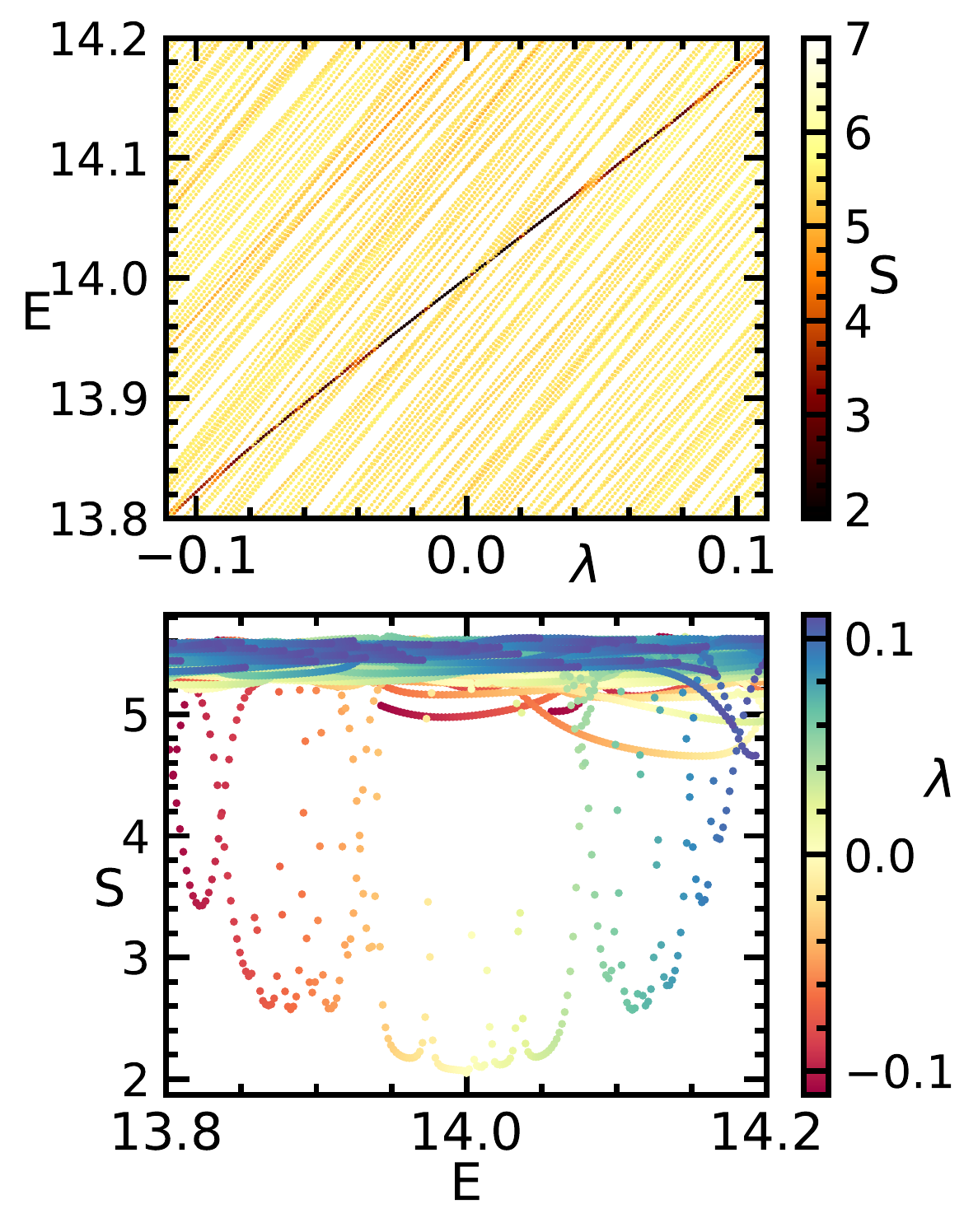}
\end{center}
\caption{
Bipartite entanglement entropy as a function of the energy of the eigenstates for different values of $\lambda$ ($L=14$). The eigenstates are in the translation- and reflection-invariant sector.
}
\label{fig:SvsEpert}
\end{figure}

Next, we consider the ladder case, and a perturbation of the type $H_\lambda=\lambda\sum_{i=0}^{L-1}(n_{i,1}n_{i+1,2}+n_{i,2}n_{i+1,1})$, with $n_{i,j}=\cd{i,j}c^{\mathstrut}_{i,j}$. The scars we construct for $\lambda=0$ are not exact eigenstates for $\lambda\neq 0$. We perform a numerical analysis following a previous study of perturbations in constrained spin chains~\cite{Lin_stability2019}.

In Fig.~\ref{fig:SvsEpert} we plot the bipartite entanglement entropy as a function of the energy for different values of $\lambda$. For $\lambda=0$ there is a single scar in the half-filling case in the translation- and reflection-invariant sector. For some values of $\lambda$, this scar hybridizes strongly with the continuum of states belonging to the thermal cloud, but small values of the entropy persist in a large region of $\lambda$ (large with respect to average gap at this energy density), excluding the aforementioned points. As is clear from Fig.~\ref{fig:SvsEpert}, the scarred state undergoes a large number of level crossing as $\lambda$ is varied but its entanglement entropy remains anomalously small. The phenomenology is extremely similar to the case of constrained spin-models, and, while system sizes here are insufficient to draw conclusions that hold in the thermodynamic limit, we can still observe the same type of resilience of scarred features at finite size.

In terms of physical implementations, the models we discussed have been partly addressed in works related to fermionic Rydberg-dressed atoms (at least, for the case of ladders). We note, however, that in terms of experimental signatures the connection to experiments requires some extra care with respect to other spin models. In order to have long-time coherent oscillations, like the ones observed in \cite{Bernien2017}, a set of equally-spaced energy eigenstates is needed. In our case, this could be achieved by adding a chemical potential, which shifts the scars according to the number of particles. However, to detect the oscillations, one should be able to prepare an initial state in a superposition with different numbers of particles. While this might be possible for spin systems (a similar mechanism is used, for example, in \cite{Schecter2019}), it is not feasible for number-conserving fermionic particles. A more direct experimental proof of the existence of scars would be obtained using the scar itself as initial state of the dynamics: every observable should remain approximately constant in time.

We comment on the connection of the eigenstates discussed above with the Shiraishi-Mori construction for embedding ETH-violating states in an otherwise ergodic spectrum~\cite{Shiraishi2017}. The construction consists of local projectors $P_j$ and a subspace $\mathcal T$ of the Hilbert space satisfying $P_j \mathcal T=0$. Then the Hamiltonian
\begin{equation}\label{eq:shmo}
    H=\sum_j P_j h_j P_j + H' \hspace{1cm} [H', P_j]=0
\end{equation}
has candidate scarred eigenstates in the subspace $\mathcal T$.
It can be shown that the Hamiltonian~(\ref{eq:Ham}) can be recast in the form of Eq.~(\ref{eq:shmo}) with $\mathcal{T}$ being the subspace with a single scar state. We examine, for instance, the scar $\ket{\psi_{A,e}}$ in Eq.~(\ref{eq:exstA}). To prove the construction, we define $P_j$ as a local projector acting on the $j$-th plaquette and on its neighbours,
\begin{equation}
    P_j=1-\ket{j_{A,e}}\bra{j_{A,e}},
\end{equation}
\begin{equation}
    \ket{j_{A,e}}=\prod_{i=j-1}^{j+1}\frac{1}{\mathcal{N}_{i,A}}\left(d^\dagger_{2i,1}-d^\dagger_{2i+1, 2}\right)\ket{0}_i.
\end{equation}
This projector annihilates the state that has a single fermion in a superposition on the $A$ diagonal in each of the plaquettes considered (as in the state $\ket{\psi_{A,e}}$) and acts trivially on the other states. We find that the term $V$ commutes with the projectors $P_j$ and corresponds to $H'$ in the Shiraishi-Mori construction. The hopping terms, on the other hand, need some further manipulation. We define $h_j$ as made of two parts: (i) the sum of the hopping terms in the $j$-th plaquette, (ii) the sum of the hopping terms between the $j$-th plaquette and its neighbors (with a factor $1/2$). With this definition, we see that $h_j=P_j h_j P_j$ and $H_0=\sum_j h_j$, resulting in the desired form of Eq.~(\ref{eq:shmo}). This construction can be applied to the other scars, and to the case of higher dimensionality.
Each scar represents an isolated embedded subspace, and hence its entanglement entropy does not scale with $L$.

\section{Conclusions and outlook} We have shown that $N=2$ supersymmetric lattice models display weak-ergodicity breaking in the form of scarred eigenstates in any $D$-dimensional hypercubic lattice. 
SUSY is not a sufficient ingredient for quantum scars in $D>1$: for instance, even within the model we consider, the spectrum at low-filling does not feature ergodicity breaking. Instead, we find important to emphasize that the results reported here underline that insights from quantum field theory - in our case, provided by the Coleman-Mandula theorem - can provide a very simple tool to easily diagnose conditions that favor quantum scarring, that is complementary to other approaches based on exact lattice solutions, that are typically applicable to single models~\cite{LinMPS2019,Moudgalya2018,Moudgalya2018a,Mark2020}. It is important to stress that it would not be sufficient for a lattice model to recover SUSY as a low-energy symmetry, since the phenomena we are concerned with require finite-energy-density above the ground state. Due to the fact that formulating explicit supersymmetric theories on the lattice is challenging, it stands as an open quest to determine if there exists additional features that, in combination with SUSY, can guarantee the appearance of quantum scars in given lattice models. To resolve such questions, it would thus be important to formulate lattice models with richer supersymmetric structures, and investigate their SUSY-specific dynamical effects~\cite{Cubero_2016}.

\begin{acknowledgments}
We acknowledge several useful discussions with D. Abanin, P. Fendley, W. Ho, M. Lukin, K. Papadodimas, and H. Pichler, and thank A. Gambassi, A. Lerose, and P. P. Mazza for collaboration on a related work. This work is partly supported by the ERC under grant number 758329 (AGEnTh), by the Quantera programme QTFLAG, and has received funding from the European Union's Horizon 2020 research and innovation programme under grant agreement No 817482 (Pasquans). This work has been carried out within the activities of TQT. 

{\it Note added. --} During completion of this work, a preprint appeared~\cite{lin2020quantum} showing how the bosonic counterpart of some of the states we discuss are also present in the 2D PXP model in the middle of the spectrum, where scarring was already observed in numerical simulations~\cite{michailidis2020stabilizing}.
\end{acknowledgments}

\bibliographystyle{unsrtaipauth4-1.bst}
\bibliography{bib}

\end{document}